observed in the δ[18]O and δD of rainfall over peninsular and north eastern India[5,6]. Besides these seasonal differences, the isotopic compositions of vapour show short term variations (time scales of a few hours to a few days), mainly attributable to cyclonic activity occurring over the sea. Isotopic compositions of water vapour show the lowest values during monsoon depressions and tropical cyclone activity. Such sharp [18]O depletions in surface vapour and rain are generally observed in the tropics in large scale organized convective systems[7–9].

We compared the measured δ[18]O of vapor with Craig and Gordon[10] model, which estimates δ[18]O and δD from surface water isotopic composition, sea surface temperature (SST) and relative humidity. δ[18]O of water vapour appeared to be close to the model estimates during non-rainy days of ISM. However, during rain events associated with monsoon depression, vapour δ[18]O values were significantly lower than the model predicted values. This may be due to the rain-vapour interaction and convective downdrafts in the upstream region. The model assumes that local oceanic evaporation is the only source of vapour (i.e., 'local closure assumption'), while these secondary processes make more [18]O depleted surface vapour. During NEM season the baseline of vapour δ[18]O is well below that of Craig and Gordon model estimation, with the lowest values associated with cyclonic activity. The more negative δ[18]O of the vapour during NEM may possibly be due to multiple factors such as i) change ocean surface conditions (relative humidity and SST), ii) δ[18]O depletion in the surface sea water due to the increased river runoff by the end of ISM season[6,11] iii) shift in circulation pattern and iv) change in the weather systems (i.e., monsoon storm/depressions vs. tropical cyclones).

According to the Craig and Gordon model, a negative relation is expected between *d*-excess (δD-8* δ[18]O) and the relative humidity. This relation arises from the diffusive transport of water isotopologues in air, during evaporation from the ocean. During condensation, which occurs usually under isotopic equilibrium condition, the rain preserves the *d*-excess values of its source vapour. Hence the *d*-excess in ice cores is known to preserve the source vapour signature. This relation was observed earlier over the Southern Indian Ocean[2] and the Atlantic Ocean[12]. We also observe such a relation over BoB, but it appears to be weaker during the ISM, and dominant during NEM.

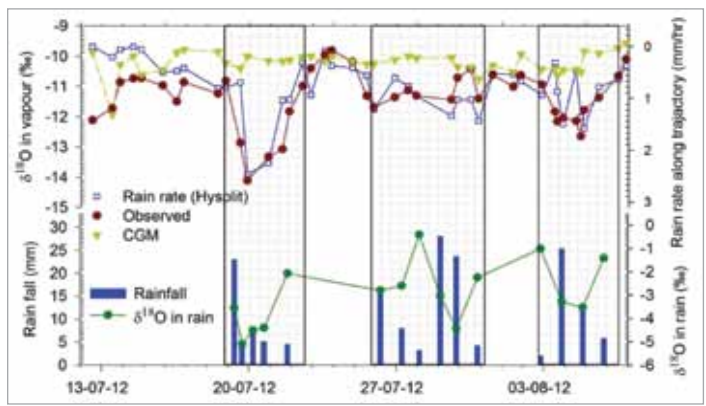

*Fig. 2. Comparison of observed δ[18]O of vapour (filled circles) with Craig and Gordon model results (CGM, inverted triangles). The shaded area shows three rain spells that occurred during sampling. The rain rate (open square) plotted in the upper panel is the average rain rate along the 24h air parcel back trajectory[13]. The accumulated rain amount collected during the cruise (vertical bars) and its δ[18]O of rain (filled circle) are also shown in bottom panel.*

# The Transporters - Physics: Ocean, Atmosphere

## A short perspective on the Mascarene High and the abnormal Indian Monsoon during 2015

R. Krishnan, Bhupendra Singh, R. Vellore,
M. Mujumdar, P. Swapna, Ayantika Choudhury
Manmeet Singh, B. Preethi and M. Rajeevan
*Indian Institute of Tropical Meteorology,
Pune 411008*

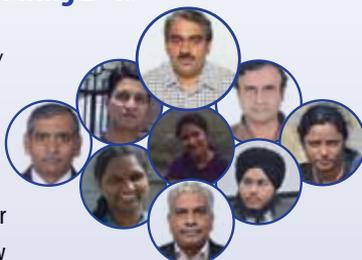

The initiation of the Indian summer monsoon circulation during late May / early June arises through large-scale land-sea thermal contrast and setting up of negative pressure gradient between the Monsoon Trough over the Indo-Gangetic plains and the Mascarene High over the subtropical Indian Ocean. The meridional pressure gradient together with the Earth's rotation (Coriolis force) creates the summer monsoon cross-equatorial flow, while feedbacks between moisture-laden winds and latent heat release from precipitating systems maintain the monsoon circulation during the June-September (JJAS) rainy season (Krishnamurti and Surgi, 1987). This simplified view of the Indian monsoon is a useful starting point to draw insights into the variability of the large-scale monsoon circulation.

In this article, we provide a short perspective on the 2015 Indian summer monsoon season – a period characterized by anomalously weak large-scale monsoon winds and deficient rains over the Indian subcontinent. With pronounced decrease of rains over the west coast and areas in peninsular and north India during 2015, the resultant deficit in the All India monsoon rainfall was about 15% of the long-term average and was accompanied by a significant weakening of the large-scale monsoon cross-equatorial flow (Ref: Fig.s1, 2). The weakened monsoon circulation during 2015 is clearly evident from the low-level easterly anomalies over the Arabian Sea and the Indian region. Further, one can notice a clear weakening of the Mascarene anticyclone as evidenced from the circulation anomalies over the subtropical Indian Ocean to the east of Madagascar and southern Africa (Fig.2b).

The weak Indian monsoon of 2015 happened to coincide with the evolution of another major climatic phenomenon – a relatively strong El Nino event in the Pacific. The westerly anomalies over the tropical Pacific shown in Figure.2b illustrate a weakening of the easterly trades seen during El Nino conditions. The El Nino related variations during 2015 are clearly borne out in the spatial patterns of sea tropical central-eastern Pacific and negative anomalies in the west Pacific depict the classical El Nino pattern. Also notice that the warm SST anomalies in the east extended far northward into the extra-tropics into the west coast of the US and northeast Pacific during 2015. The negative (positive) SLP anomalies over the central-eastern (western) Pacific basically correspond to a weakening of the Walker circulation (Fig.3). The warm SST anomalies in the Northern Indian Ocean and western Arabian Sea are typically associated with weakened monsoon winds, decreased evaporation and reduced upwelling along the Somali and Oman coasts (ex., Rao, 1987, Ramesh and Krishnan, 2005, Scott et al. 2009). It is also noteworthy to mention here that the Pacific SST anomalies in 2015 had resemblance with those during the strong El Nino of 1997 (not shown). Interestingly, it must be pointed out that the strong El Nino of 1997 had little impact on the Indian monsoon which actually turned out to be a near-normal monsoon rainy season (http://www.tropmet.res.in).

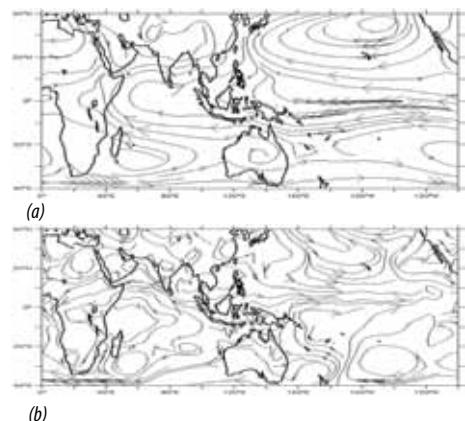

Fig.2. Mean and anomaly map of 850 hPa streamlines for the JJAS season (a) Climatological mean (b) Anomaly for 2015. The wind dataset is from NCEP reanalysis for the period (1981-2014)
Ref:http://www.esrl.noaa.gov/psd/data/gridded/data.ncep.reanalysis.derived.html

Although notwithstanding the likely El Nino impact on the Indian monsoon in 2015, the anomalous conditions in the southern Indian Ocean during that period lead us to think about possible non-El Nino factors that might have significantly influenced the large-scale monsoon circulation. For example the Indian Ocean SLP anomalies during 2015 show a weakened Mascarene High along with warm SST anomalies east of Madagascar in the southern Indian Ocean (see Fig.3). These anomalous features during 2015 are suggestive of a likely influential role of the weak Mascarene High on the large-scale monsoon flow and rains over India. Several studies have drawn attention to the role of the southern Indian Ocean SST variability in influencing the Southern Hemispheric climate during the Austral summer (eg., Behera and Yamagata, 2001; Chiodi and Harrison, 2007; Wang, 2010, Morioka et al., 2014, Ohishi et al. 2015). However, aspects of ocean/atmospheric coupled processes in the southern Indian Ocean have received less attention for the Austral winter (or boreal summer) season (Boschat et al. 2011). For example, there is lack of understanding on the basic mechanisms that give rise to persistent SST and SLP anomalies in the southern Indian Ocean during the boreal summer and how they interact with the large-scale monsoon circulation. The boreal summertime teleconnection between the Indian monsoon and the southern Indian Ocean SST/SLP variability assumes greater significance especially given the rapid rate of Indian Ocean warming during recent decades (Copsey et al. 2006). To summarize, a comprehensive understanding of the connection among the Indian Monsoon and the variability of the Mascarene High and the southern Indian Ocean SST is an important scientific problem and warrants further investigations.

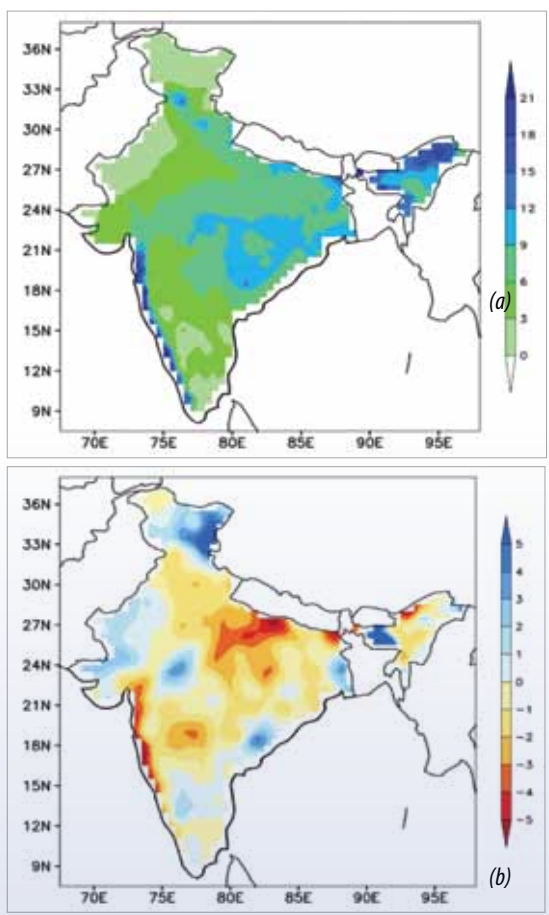

Fig.1. (a) Spatial map of long-term (1979-2015) climatological mean rainfall (mm day-1) for the JJAS season (b) Rainfall anomaly for JJAS 2015. The rainfall dataset is based on the India Meteorological Department (IMD) dataset.
Ref: http://www.imd.gov.in/section/nhac/dynamic/Monsoon_frame.htm

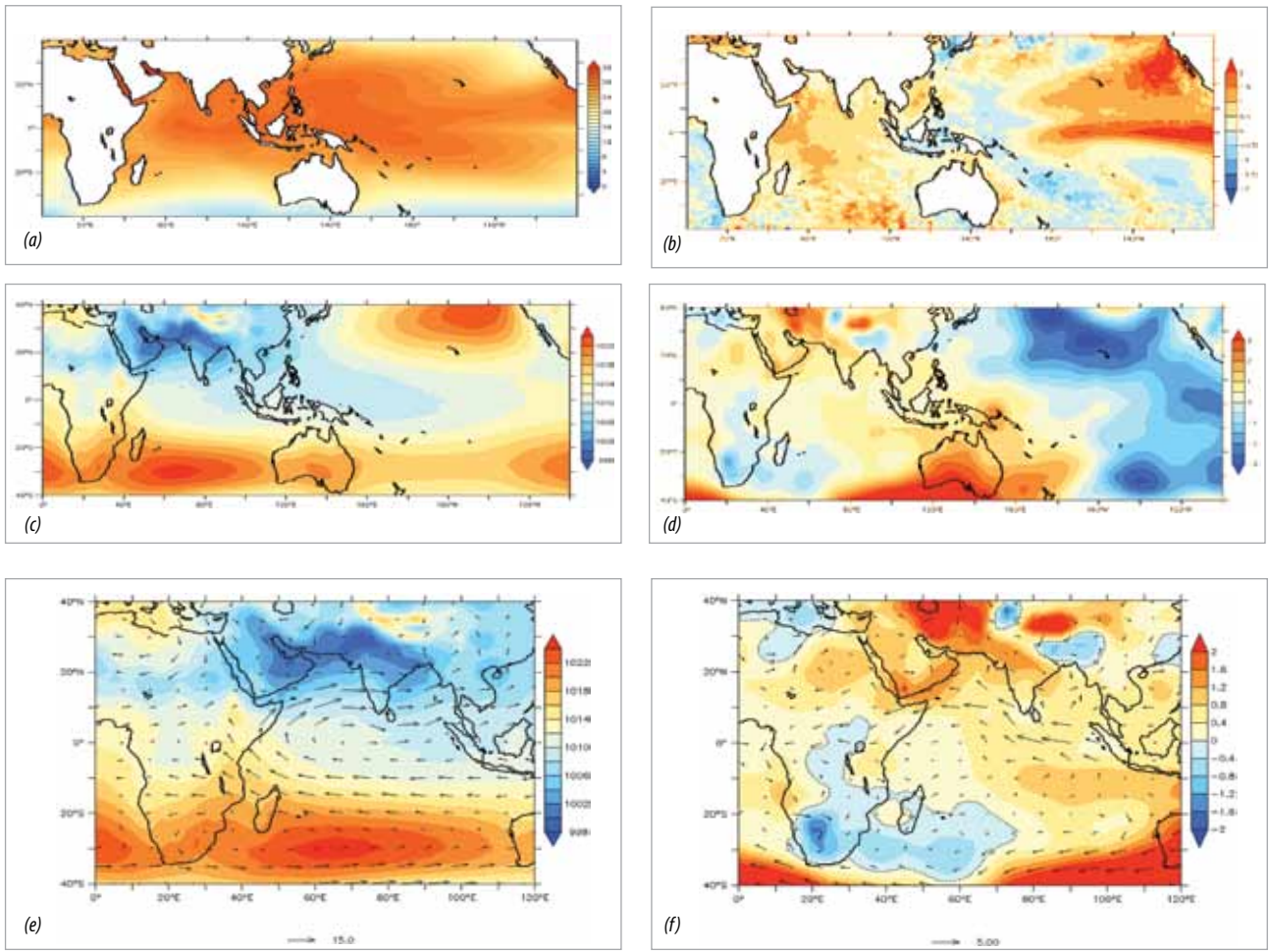

*Fig.3. Mean and anomaly maps of SST (°C), SLP (hPa) and winds at 850 hPa for the JJAS season (a) Mean SST (b) SST anomaly during 2015 (c) Mean SLP (d) SLP anomaly during 2015 (e) Mean winds and SLP over the Indian Ocean and monsoon region (f) Wind and SLP anomaly during 2015. The mean fields are for the period (1981-2014) based on the datasets: OISST http://www.esrl.noaa.gov/psd/data/gridded/data.noaa.oisst.v2.highres.html and NCEP reanalysis http://www.esrl.noaa.gov/psd/data/gridded/data.ncep.reanalysis.derived.html*